\documentclass[]{aa}

\usepackage{amsmath}
\usepackage{amssymb}
\usepackage{url}
\usepackage{graphicx}
\usepackage{times}
\usepackage[utf8]{inputenc}
\usepackage{xspace}
\usepackage{color}
\usepackage{multirow}
\usepackage{multicol}
\usepackage{float}

\bibpunct{(}{)}{;}{a}{}{,}

\usepackage{etoolbox}
\makeatletter
\patchcmd\@combinedblfloats{\box\@outputbox}{\unvbox\@outputbox}{}{%
  \errmessage{\noexpand\@combinedblfloats could not be patched}%
}%
\makeatother

\def\apjl{ApJL}
\def\apjs{ApJSS}

\def\aap{A\&A}

\def\eg{e.g.\xspace}

\def\grbepeak{$E_{\rm peak} = 185 \pm 62\,\rm keV$\xspace}

\def\E0fit{$E_0=(3.1^{+2.1}_{-1.3})\times 10^{49}\,{\rm erg}$}

\def\R0fit{$R_0=54^{+50}_{-25}\,{\rm km}$}
\def\T0fit{$T_0=(2.6^{+1.5}_{-1.0})\times 10^{11}\,{\rm K}$}

\def\liso{$L_{\rm iso}$}
\def\eiso{$E_{\rm iso}$}

\title{The on-axis view of GRB 170817A}
\titlerunning{on-axis view of GRB 170817A}
\author{O.~S.~Salafia\inst{\ref{oab.me},\ref{infn.mib}}, G.~Ghirlanda\inst{\ref{oab.me},\ref{infn.mib}}, S.~Ascenzi\inst{\ref{gssi}} and G.~Ghisellini\inst{\ref{oab.me}}}

\institute{INAF -- Osservatorio Astronomico di Brera, via E. Bianchi 46, I-23807 Merate (LC), Italy\label{oab.me} \and INFN -- Sezione di Milano-Bicocca, Piazza della Scienza 3, I-20126 Milano (MI), Italy\label{infn.mib} \and Gran Sasso Science Institute, viale F. Crispi 7, I-67100 L'Aquila (AQ), Italy\label{gssi}}

\authorrunning{Salafia, Ghirlanda, Ascenzi \& Ghisellini}

\date{Received xxx / Accepted: xxx}

\abstract{The peculiar short gamma-ray burst GRB 170817A has been firmly associated to the gravitational wave event GW170817, which has been unaninmously interpreted as due to the coalescence of a double neutron star binary. The unprecedented behaviour of the non-thermal afterglow led to a debate about its nature, which was eventually settled by high-resolution VLBI observations, which strongly support the off-axis structured jet scenario. Using information on the jet structure derived from multi-wavelength fitting of the afterglow emission and of the apparent VLBI image centroid motion, we compute the appearance of a GRB 170817A-like jet as seen by an on-axis observer and we compare it to the previously observed population of SGRB afterglows and prompt emission events. We find that the intrinsic properties of the GRB 170817A jet are representative of a typical event in the SGRB population, hinting at a quasi-universal jet structure. The diversity in the SGRB afterglow population could therefore be ascribed in large part to extrinsic (redshift, density of the surrounding medium, viewing angle) rather than intrinsic properties. Although more uncertain, the comparison can be extended to the prompt emission properties, leading to similar conclusions.
}


\begin{document}

\maketitle

\section{Introduction}

The discovery of the short gamma-ray burst (SGRB) GRB 170817A by Fermi/GBM \citep{Goldstein2017} and INTEGRAL/SPI-ACS \citep{Savchenko2017} in concert with the Advanced LIGO/Virgo gravitational wave (GW) event GW170817 produced by the merger of two neutron stars \citep{Abbott2017} unleashed a new era in multi-messenger astronomy \citep{Abbott2017a}.  The $\gamma$-ray emission, delayed by 1.734$\pm$0.054 s \citep{Abbott2017a} with respect to the GW event chirp, consists of a short burst of duration $T_\mathrm{90}\sim 2.0 \pm 0.5\,\mathrm{s}$, with evidence of a sub-structure characterised by a short ($\sim 0.5\,\mathrm{s}$) hard spike followed by a longer and softer tail \citep{Goldstein2017,Zhang2017}. The 10--1000 keV peak flux ($3.7 \pm 0.9\,\mathrm{ph\,cm^{-2}\,s^{-1}}$) and fluence ($2.8 \pm 0.2 \times 10^{-7}\,\mathrm{erg\,cm^{-2}}$), combined with the nearby distance of the source, make GRB 170817A several orders of magnitude less luminous and less energetic than typical short GRBs \citep{Abbott2017a}. With the spectral parameters provided by \citet{Goldstein2017} and assuming a luminosity distance $d_\mathrm{L}=41\,\mathrm{Mpc}$ \citep{Coulter2017}, the isotropic equivalent peak luminosity is $L_\mathrm{iso}=(1.4\pm0.5) 10^{47}\,\mathrm{erg\,s^{-1}}$ and the isotropic equivalent energy is $E_\mathrm{iso}=(5.6\pm 0.4)\times 10^{46}\,\mathrm{erg}$. The peak of the best fit $\nu F(\nu)$ spectrum during the initial hard spike is \grbepeak \citep{Goldstein2017}\footnote{Similar results were obtained by an independent analysis by \cite{Zhang2017}}.

The discovery, about 12 hours after the GBM trigger, of the associated optical transient AT2017gfo \citep{Coulter2017,Valenti2017} triggered an intense ultraviolet, optical and infrared (UVOIR) follow up campaign \citep[e.g.][]{Andreoni2017,Arcavi2017,Chornock2017,Covino2017,Diaz2017,Drout2017,Evans2017,Hallinan2017,Pian2017,Pozanenko2017,Smartt2017,Utsumi2017} 
-- whose results have been collected and homogenized in \cite{Villar2017} -- which characterised with unprecedented details the emission and color evolution of the first ever spectroscopically confirmed \citep[\eg][]{Pian2017,Cowperthwaite2017,Gall2017,Kilpatrick2017,McCully2017,Nicholl2017,Smartt2017} kilonova (KN), i.e.~nuclear-decay-powered emission from the expanding ejecta produced in the double neutron star (NS-NS) merger \citep{Li1998,Metzger2011,Metzger2016}. 

Another emission component was detected in X-rays $\sim 9$ days after the merger by \textit{Chandra} \citep{Troja2017,Margutti2017,Haggard2017} and one week later in the radio by the Karl G. Jansky Very Large Array (JVLA) \citep{Hallinan2017}. Subsequent monitoring of the flux density in several bands \citep[e.g.][]{Margutti2017,Troja2017,Mooley2017,Lamb2018a,Lyman2018,Margutti2018,Troja2018,DAvanzo2018,Dobie2018,Resmi2018} established the non-thermal nature of this emission component, but the community could not agree on its interpretation until more than 200 days post-merger. High-resolution very long baseline interferometry (VLBI) observations \citep{Mooley2018,Ghirlanda2019} eventually provided strong support to the interpretation of this emission component as being synchrotron emission from the forward shock caused by a relativistic, narrow jet sweeping the interstellar medium (ISM). Within this scenario, the evolution of the flux density requires the jet to feature an angular ``structure'', namely a narrow core (of half-opening angle $\theta_\mathrm{c}\sim 3^\circ$) with an approximately uniform distribution of kinetic energy per unit solid angle $dE/d\Omega(\theta)$ and Lorentz factor $\Gamma(\theta)$ (here $\theta$ is the angle from the jet axis), surrounded by ``wings'' where both the kinetic energy density and Lorentz factor decrease steeply as a function of $\theta$ \citep[e.g.][]{Lazzati2017,Lyman2018,Troja2018,Lamb2018a,Resmi2018,Margutti2018,Hotokezaka2018a,Ghirlanda2019,Ziaeepour2019}.

Despite several uncertainties in the modelling and the large number of  parameters, quite remarkably the analyses carried out by different groups with diverse methodologies largely agree about the general features of the jet structure (see Figure~S6 in \citealt{Ghirlanda2019} for a visual comparison). In this work we address whether a jet with this structure, when observed on-axis, does resemble the known SGRB population, and to what extent. We investigate this question for both the afterglow emission, by comparing the on-axis emission of a GRB 170817A-like jet to that of a large sample of SGRBs collected by \citet{Fong2015}, and for the prompt emission, using an updated version of the SBAT4 sample \citep{DAvanzo2014} as comparison. These comparisons provide suggestive hints towards the possibility that most SGRBs possess very similar jets, implying that a large part of their diversity could be ascribed to extrinsic differences (viewing angle, ISM density, redshift).

\section{Afterglow model}

We describe here in brief the afterglow model we use to compute the light curves presented in this work. The model, based on standard GRB afterglow concepts, has been used previously in \citet{DAvanzo2018} and \citet{Ghirlanda2019}, and is under many aspects similar to those of \citet{Lamb2017} and \citet{Gill2018}. 

\subsection{Dynamics}

Let us define a spherical coordinate system whose $z$ axis is aligned with the jet axis, and let us call $\theta$ and $\phi$ the latitudinal and azimuthal angles respectively. The observer lies on the $(z,y)$ plane, her line of sight forming an angle $\theta_\mathrm{v}$ with the jet axis. 
Let us call ``annulus'' the part of the jet comprised between $\theta$ and $\theta+d\theta$. We consider each annulus independently, ignoring energy transport between adjacent annuli (i.e.~we neglect the jet lateral expansion, which only affects the late-time behaviour of the light curves, and has no effect on our conclusions -- we will present a model of lateral expansion in a subsequent work). The initial kinetic energy per unit solid angle in each annulus is $dE/d\Omega$, and its initial Lorentz factor is $\Gamma(0,\theta)$.
As the annulus expands, it sweeps the ISM producing a shock. At a given radius $R$, the swept mass per unit solid angle amounts to
\begin{equation}
    \mu(R) = \frac{R^3}{3} n m_\mathrm{p}
\end{equation}
where $n$ is the ISM number density and $m_\mathrm{p}$ is the proton mass (i.e.~we assume the ISM is made of pure hydrogen).
We compute the dynamics of the shocked material enforcing energy conservation, following \citet{Panaitescu2000} and \citet{Granot2003}. This yields the Lorentz factor of the shocked material\footnote{the material behind the shock actually features a Lorentz factor profile \citep{Blandford1976}. This energy conservation argument yields the average Lorentz factor. Since most material is concentrated right behind the shock, though, the result is close to the Lorentz factor just upstream of the shock.}, namely
\begin{equation}
    \Gamma(R,\theta) = \frac{\mu_\mathrm{0}}{2 \mu}\left[\sqrt{1+\frac{4 \mu(c^{-2}dE/d\Omega+\mu+\mu_\mathrm{0})}{\mu_\mathrm{0}^2}}-1\right]
\end{equation}
where $\mu_\mathrm{0}(\theta)=[dE/d\Omega(\theta)]/[\Gamma(0,\theta)c^2]$ and $c$ is the speed of light. The shocked material is concentrated in a thin layer behind the shock \citep{Blandford1976}, which we approximate as having a uniform radial density distribution within a thickness $\Delta R$, which depends on $\theta$, $R$ and $\Gamma$.
By enforcing electron number conservation, we can relate this thickness to the other quantities defining the dynamics. We compute the electron number density $n_\mathrm{s}$ of the shocked material imposing the shock-jump condition \citep{Blandford1976}
\begin{equation}
    n_\mathrm{s} = \frac{\gamma_\mathrm{ad}\Gamma+1}{\gamma_\mathrm{ad}-1}n
\end{equation}
where $\gamma_\mathrm{ad}$ is the post-shock adiabatic index. We approximate $\gamma_\mathrm{ad}$, as a function of $\Gamma$, using the fitting function by \citet{Peer2012}, which gives the correct post-shock adiabatic index under the assumption of (i) a strong shock and (ii) that the shocked particle energy distribution is dominated by a Maxwellian. This allows us to derive the thickness of the shocked layer as
\begin{equation}
    \Delta R = \frac{R(\gamma_\mathrm{ad}-1)}{3(\gamma_\mathrm{ad}\Gamma+1)\Gamma}
\end{equation}
The shocked layer grows thicker as more and more ISM matter is swept, with the forward shock moving faster than the shocked material itself. Its Lorentz factor $\Gamma_\mathrm{s}$ can be related to $\Gamma$ by \citep{Blandford1976}
\begin{equation}
    \Gamma_\mathrm{s} = (\gamma_\mathrm{ad}(\Gamma-1)+1)\sqrt{\frac{\Gamma+1}{\gamma_\mathrm{ad}(2-\gamma_\mathrm{ad})(\Gamma-1)+2}}
\end{equation}

\subsection{Equal-arrival time surfaces}
Photons emitted by the material behind the shock at a given time reach the observer at different arrival times. Let us assume, as a simplifying approximation, that all the emission comes from the shock surface, which is justified as the emitting layer is thin compared to the shock radius. The relation between the emission radii $R(\theta,\phi)$ and the arrival times $t_\mathrm{obs}$ is therefore
\begin{equation}
    t_\mathrm{obs}(R,\theta,\phi,\theta_\mathrm{v}) = (1+z)\int_0^R \frac{dR(1-\beta_\mathrm{s}\cos\alpha)}{\beta_\mathrm{s}c}
    \label{eq:tobs}
\end{equation}
where $z$ is the redshift, 
$\beta_\mathrm{s}=(1-\Gamma_\mathrm{s}^{-2})^{1/2}$, $\cos\alpha = \cos\theta\cos\theta_\mathrm{v}+\sin\theta\sin\phi\sin\theta_\mathrm{v}$ (given our assumed geometrical setting) and we choose the integration constant so to set $t_\mathrm{obs}=0$ for a photon emitted when the shock radius was $R=0$. We write the shock surface brightness as
$I_\mathrm{\nu}(\nu,R,\theta,\phi)=\delta^3 \Delta R' j'_{\nu'}(\nu/\delta)$, where $\delta(R,\theta,\phi,\theta_\mathrm{v})=\Gamma(R,\theta)^{-1}[1-\beta(R,\theta)\cos\alpha]^{-1}$ is the Doppler factor (note that it is computed using the shocked material velocity $\beta=(1-\Gamma^{-2})^{1/2}$, as opposed to the shock velocity), $\Delta R' = \Gamma(R,\theta)\Delta R$ and $j'_{\nu'}$ is the comoving emissivity, which we assume to be due to synhcrotron emission as detailed in the following subsection. In order to compute the light curves, we set up a grid with $N_\theta$ latitudinal divisions $\theta_\mathrm{i}$, equally spaced in the logarithm so that $10^{-4}\leq\theta_\mathrm{i}\leq \pi/2$, and $N_\mathrm{\phi}$ azimuthal divisions $\phi_\mathrm{j}$ with $-\pi/2\leq \phi_\mathrm{j} \leq \pi/2$ (i.e.~we only cover half of the jet, to exploit the symmetry of the jet image under reflection across the $(z,y)$ plane). We compute the surface brightness at each point of the grid, at radii corresponding to a given arrival time $t_\mathrm{obs}$, and we finally integrate over the grid to get the flux density at that time, i.e.
\begin{equation}
    F_\mathrm{\nu}(\nu,t_\mathrm{obs}) = 2\times\frac{1+z}{d_\mathrm{L}^2}\int_0^{1}d\cos\theta\int_{-\pi/2}^{\pi/2}d\phi \,\, R^2 I_\mathrm{\nu}((1+z)\nu,R)
\end{equation}
where $R=R(\theta,\phi,t_\mathrm{obs})$, i.e.~the equal-arrival-time surface, is obtained by inverting Eq.~\ref{eq:tobs}, and the factor $2$ is to recover the flux from the whole jet, as we only compute it on half of the solid angle, as explained above. To compute the flux from the counter-jet, which we assume to possess the same properties as the jet, we follow the same procedure, but setting $\theta_\mathrm{v}\to \pi + \theta_\mathrm{v}$.

\subsection{Radiation}
We model the emission from the shocked material (assuming only a forward shock is present) in a way similar to \citet{Sari1998} and \citet{Panaitescu2000}. We assume shocked ISM electrons to be accelerated into a power law distribution in Lorentz factor, namely
\begin{equation}
    \frac{dn_\mathrm{s}}{d\gamma}\propto \gamma^{-p}
\end{equation}
with $p>2$, above a minimum (``injection'') electron Lorentz factor $\gamma_\mathrm{m}$. We assume that their total energy density amounts to a fraction $\epsilon_\mathrm{e}$ of the post-shock energy density, defined as $e=(\Gamma-1)n_\mathrm{s}m_\mathrm{p}c^2$. These assumptions lead to the following definition of the injection Lorentz factor \citep{Sari1998}
\begin{equation}
    \gamma_\mathrm{m}= \max\left[1,\frac{p-2}{p-1}(\Gamma-1)\frac{m_\mathrm{p}}{m_\mathrm{e}}\right]
\end{equation}
where $m_\mathrm{e}$ is the electron rest mass, and we ensure $\gamma_\mathrm{m}\geq 1$. We assume the magnetic field upstream of the shock to be amplified by small-scale instabilities to an energy density $B^2/8\pi$ equal to a fraction $\epsilon_\mathrm{B}$ of the post-shock total energy density $e$. With these assumptions, we can compute the comoving synchrotron emissivity of electrons behind the shock at the peak of their spectrum (from e.g.~\citealt{VanEerten2011}, with the modification described in \citealt{Sironi2013} to account for the ``deep newtonian'' regime) 
\begin{equation}
    j'_{\nu',\mathrm{max}} \approx 0.66\frac{q_\mathrm{e}^3}{m_\mathrm{e}^2 c^4}\frac{p-2}{3p-1}\frac{B \epsilon_\mathrm{e} e}{\gamma_\mathrm{m}}
\end{equation}
where $q_\mathrm{e}$ is the electron charge. We then write $I_\nu (\nu) = \delta^3 \Delta R' j'_\mathrm{\nu',max}\Sigma(\nu/\delta)$, where $\Sigma(\nu')$ is the normalized spectral shape, which can be approximated by a series of power laws (we implement all spectral orderings as listed in \citealt{Granot2002}). The power law branches connect at the break frequencies $\nu_\mathrm{m}$, $\nu_\mathrm{c}$, $\nu_\mathrm{a}$ and $\nu_\mathrm{ac}$ defined as follows: $\nu_\mathrm{m}$ is the synchrotron frequency corresponding to the injection Lorentz factor, namely $\nu_\mathrm{m}=\gamma_m^2 q_\mathrm{e}B/2\pi m_\mathrm{e}c$. Similarly, $\nu_\mathrm{c}$ is the synchrotron frequency corresponding to the Lorentz factor $\gamma_\mathrm{c}$ above which electrons cool faster than the expansion dynamical timescale \citep[i.e.~the timescale over which new electrons are accelerated and injected into the shocked region --][]{Sari1998}, namely
\begin{equation}
    \gamma_\mathrm{c}=\frac{6\pi m_\mathrm{e}c^2\Gamma\beta}{\sigma_\mathrm{T}B^2 R}
\end{equation}
where $\sigma_\mathrm{T}$ is the Thomson cross section. The frequencies $\nu_\mathrm{a}$ and $\nu_\mathrm{ac}$ are related to synchrotron self-absorption. In the slow-cooling regime ($\nu_\mathrm{m}<\nu_\mathrm{c}$), $\nu_\mathrm{a}$ marks the frequency below which the emission is self-absorbed. We compute this frequency following \citet{Panaitescu2000}. In the fast-cooling regime ($\nu_\mathrm{m}>\nu_\mathrm{c}$), an additional self-absorbed regime exists between $\nu_\mathrm{a}$ and $\nu_\mathrm{ac}$, due to the inhomogeneous distribution of electrons at different cooling stages: we compute the latter frequency following \citet{Granot2000}. All frequencies are computed in the comoving frame, and then transformed to the observer frame according to $\nu=\delta\, \nu'/(1+z)$. The slopes of the power law branches of $\Sigma(\nu')$ for all spectral orderings are given in \citet{Granot2002}.

\section{Comparison with known SGRB afterglows}

\begin{figure*}
    \centering
    \includegraphics[width=\textwidth]{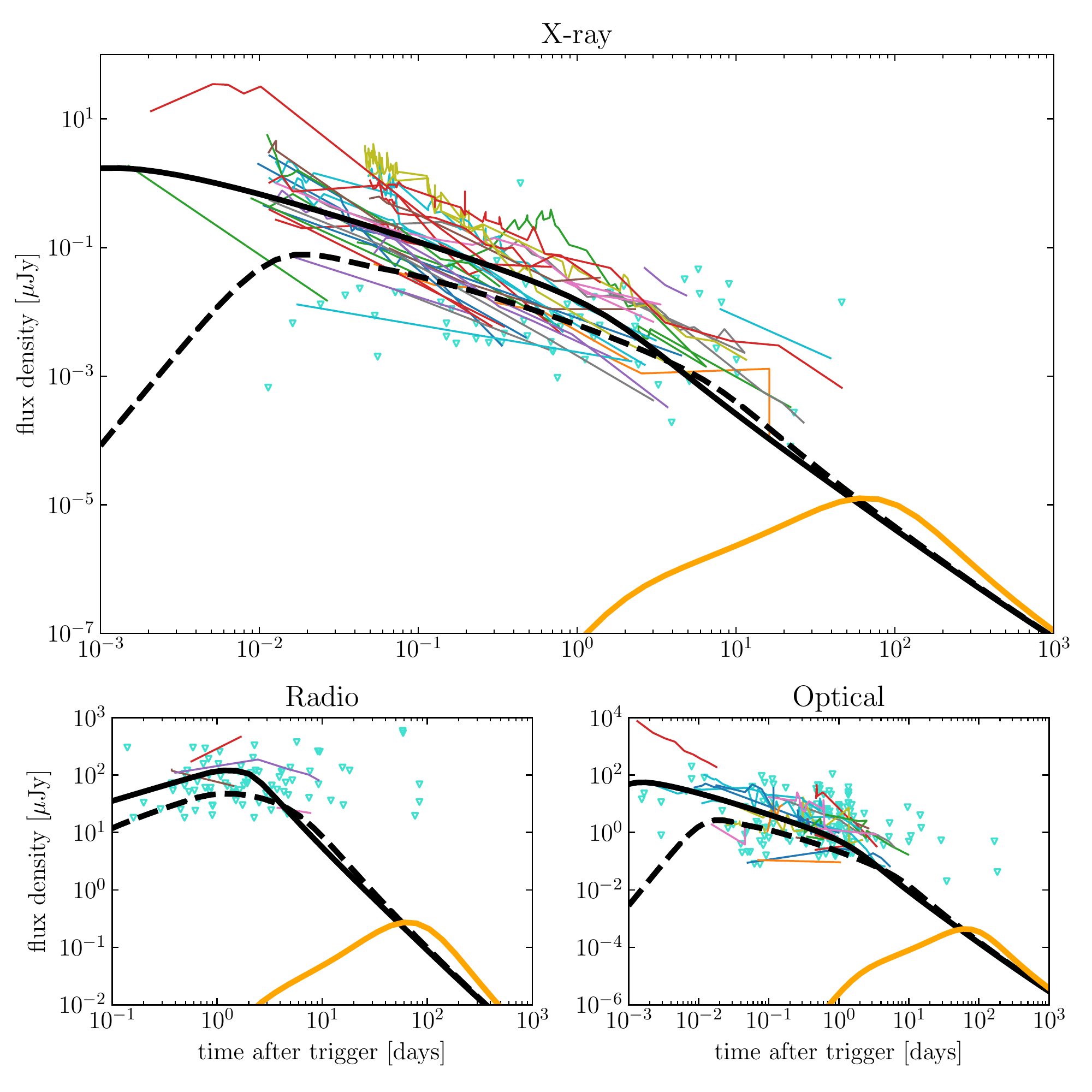}
    \caption{Afterglow of a GRB 170817A-like jet compared to the archival SGRB data from \citet{Fong2015}. The redshift and ISM density have been set equal to the medians of the \citet{Fong2015} sample, namely $z=0.46$ and $n=5\times 10^{-3}\,\mathrm{cm^{-3}}$. Each panel shows data from a different frequency range: radio ($1.4\leq \nu/\mathrm{GHz}\leq 93$, bottom left-hand panel), UVOIR (mostly r filter $\approx 4.8\times 10^{14}\,\mathrm{Hz}$, bottom right-hand panel) and X-ray (1 keV $\approx 2.4\times 10^{17}\,\mathrm{Hz}$, top panel). Upper limits are shown by empty turquoise downward triangles, while detections are shown by thin lines of different colours, each connecting data from a single SGRB. In each panel, thick solid black lines represent the GRB 170817-like jet light curve as seen on-axis, dashed black lines show the light curve for an observer at the border of the core, while solid orange lines show the light curve for viewing angle $\theta_\mathrm{v}=15^\circ$ which corresponds to the best fit value for GRB 1708017A as given in \citet{Ghirlanda2019}. The model light curves are computed for observer frequencies $\nu_\mathrm{obs}=6\,\mathrm{GHz}$ (radio), $4.8\times 10^{14}\,\mathrm{Hz}$ (optical) and $2.4\times 10^{17}\,\mathrm{Hz}$ (X-ray).}
    \label{fig:afterglows}
\end{figure*}

\citet{Fong2015} collected 103 short-duration GRBs with measured afterglow emission, either in the X-ray (71) UVOIR (30) and/or radio band (4). For a sub-sample of 32 events with well-sampled light curves, the modelling of their broadband emission with the standard afterglow model provided estimates of the isotropic-equivalent kinetic energy and circum-burst ISM density. In particular, 80\%-90\% of these short GRBs show indications for a low density ($n<1$ cm$^{-3}$) circum-burst environment with an average value of $n\sim(3-5)\times 10^{-3}$. In Fig.~\ref{fig:afterglows} we show the data points collected by \citet{Fong2015}. The radio emission of SGRBs is detected only in a handful of events \citep{Chandra2012} and its different time evolution, with respect to the UVOIR and X-ray band emission, is typically due to self-absorption at relatively early times.

We compute the afterglow emission of a GRB 170817A-like structured jet model assuming the best fit structure from \cite{Ghirlanda2019}, defined by the angular distribution of kinetic energy density per unit solid angle
\begin{equation}
    \frac{dE}{d\Omega}(\theta) = \frac{E_\mathrm{c}/4\pi}{1+(\theta/\theta_\mathrm{c})^{s_1}}
    \label{eq:dE_dOmega}
\end{equation}
and the angle-dependent initial bulk Lorentz factor
\begin{equation}
    \Gamma(0,\theta) = 1 + \frac{\Gamma_\mathrm{c}-1}{1+(\theta/\theta_\mathrm{c})^{s_2}}
    \label{eq:Gamma_theta}
\end{equation}
where $\theta$ is the angular distance from the jet axis. The best fit parameter values are $\log(E_\mathrm{c}/\mathrm{erg})=52.4^{+0.6}_{-0.7}$, $s_\mathrm{1}=5.5^{+1.3}_{-1.4}$, $\log(\Gamma_\mathrm{c})=2.4_{-0.4}^{+0.5}$, $s_\mathrm{2}=3.5_{-1.7}^{+2.1}$, and $\theta_\mathrm{c}/\mathrm{deg}=3.4^{+1.0}_{-1.0}$ (one-sigma uncertainties). We use the central values in producing the light curves in Fig.~\ref{fig:afterglows}. 
We place the jet at the median redshift ($z=0.46$) and median ISM number density ($n=5\times 10^{-3}\,\mathrm{cm^{-3}}$) of the \citet{Fong2015} sample. The orange solid lines in the three panels of Fig.~\ref{fig:afterglows} represent the model afterglow light curves for an observer at $\theta_\mathrm{v}=15^\circ$, i.e.~the same viewing angle as GRB 170817A, while solid black lines are for an on-axis observer, and dashed black lines are for an observer just outside the core ($\theta_\mathrm{v}=\theta_\mathrm{c}$). Quite remarkably, the on-axis light curves fall right in the middle of the observed population. Intriguingly, some observed UVOIR and X-ray light curves show a decay slope more similar to that of the GRB 170817A-like jet seen just outside the core. This simple comparison provides a first hint that the GRB 170817A intrinsic jet properties could be representative of a quasi-universal structure, and that the diversity in the SGRB afterglows observed so far could be largely ascribed to extrinsic properties (i.e.~redshift, ISM density, viewing angle). During the peer-review of this work, a preprint \citep{Wu2019} was circulated, which reaches similar conclusions through a slightly different analysis, strengthening our results. 

\section{Comparison of prompt emission properties}

\subsection{The  short GRB sample}\label{sec:sample}
The faintness of the afterglow emission of short GRBs hampers their prompt localisation and follow-up, thus limiting the possibility to measure their redshift. At present, such measurement was successful only for a few dozens of events. \cite{DAvanzo2014} collected a small flux-limited sample of SGRBs detected by Swift under observing conditions favourable for the redshift measurement. The sample contains 36 SGRBs detected as of 2013 and has a redshift completeness of 45\% which increases to 70\% if only the 13 brightest short GRBs are considered. For the purposes of this work we expand this sample including a number of SGRBs with measured redshift, some of which, though, do not satisfy the selection criteria to be included in the complete sample of \cite{DAvanzo2014}. In order to compute their isotropic equivalent energy $E_{\rm iso}$, we require the $\nu F_{\nu}$ peak energy of their prompt emission spectrum ($E_\mathrm{p}$) to be measured. We therefore add 10 new SGRB to the sub-sample of the 13 brightest SGRBs of \cite{DAvanzo2014}. The redshift, peak energy and isotropic equivalent energy of the entire sample are reported in Tab.~\ref{tab1}. 

The lowest redshift in this sample is 0.122 (GRB 080905A) and the isotropic equivalent energies range from a few $10^{49}$ ergs to $\sim 10^{53}$ ergs. The $\nu F_{\nu}$ peak energies are range from 100 keV to a few MeV (see also \citealt{Ghirlanda2009}). 

In long GRBs the prompt emission peak energy $E_{\rm p}$ and the isotropic equivalent energy \eiso\  or isotropic equivalent luminosity \liso\ are correlated  \citep{Yonetoku2004,Amati2002}. Short GRBs seem to follow an $E_{\rm p}$-\eiso\ correlation similar to that of long events, but slightly displaced towards lower values of \eiso\ \citep{DAvanzo2014}.  This could be ascribed to the different durations and spectral evolution of long and short GRBs \citep{Calderone2015,Ghirlanda2015}. Fitting a relation $\log(E_{\rm  p}/100\, {\rm keV})=K+\alpha \log(E_{\rm iso}/10^{52}\, {\rm erg})$ to the sample of short GRBs reported in Tab.~\ref{tab1}, the parameters are $K=0.95\pm0.07$ and $\alpha=0.48\pm0.06$ (1$\sigma$ errors). 

While all GRBs in our sample are of short duration, this does not guarantee that their progenitor is a binary neutron star merger. Even though the duration distribution of GRBs is clearly bimodal \citep{Kouveliotou1993}, the two populations overlap significantly. The duration division line (customarily taken at $T_\mathrm{90}=2\,\mathrm{s}$), moreover, is detector-dependent \citep{Bromberg2012}. A more accurate classification needs to account also for the host galaxy properties, the location of the burst within the galaxy, the possibility to firmly rule out an associated supernova, and the statistical comparison of the prompt and afterglow properties of the burst with those of the reference population. \citealt{Zhang2009} (Z09 hereafter) carried out an accurate analysis of all these properties for a sample of putative SGRBs with measured redshift, finding that only a small subset of them (which they dubbed ``Type I Gold Sample'') could be securely classified as originating from a double neutron star merger (we marked these bursts with a $z$ superscript in our Table~\ref{tab1}, and with a cyan circle in Figure~\ref{fig:prompt}). Following a different approach, namely by modelling the probability distribution of GRBs in the hardness -- duration plane, \citealt{Bromberg2013} (B13 hereafter) proposed a way to estimate the probability $f_\mathrm{NC}$ that a given GRB is of non-collapsar origin. We mark the bursts with $f_\mathrm{NC}\geq 0.5$, as given in B13, with a $b$ superscript in our Tab.~\ref{tab1}, and with a green square in our Fig.~\ref{fig:prompt}.

\begin{table}
\caption{Short GRB rest frame energetics and prompt emission peak energy. GRB 090426 and 100816 (in italics) are suspected long GRBs (see discussion in D14)}

\begin{tabular}{lccc}
\hline
  \multicolumn{1}{c}{GRB} &
  \multicolumn{1}{c}{z} &
  \multicolumn{1}{c}{$E_{\rm p}$} &
  \multicolumn{1}{c}{\eiso} \\
 &
 &
  \multicolumn{1}{c}{[keV]} &
  \multicolumn{1}{c}{[$10^{51}$ erg]}
   \\

\hline
050509B$^{z,b}$ & 0.2248     & $82_{-80}^{+611}  $ & $ 2.4_{-1.0}^{+4.4}\times 10^{-3}$\\
050709$^z$  & 0.16       & $83_{-12}^{+18}   $ & $0.033 \pm 0.001 $ \\
050724$^z$  & 0.2570     & $110_{-45}^{+400} $ & $0.09_{-0.02}^{+0.011} $ \\
051221A     & 0.547      & $621.5  \pm 127   $ & $2.6   \pm 0.35  $ \\
060614$^z$  & 0.125      & $302_{-85}^{+214} $ & $2.4   \pm 0.4   $ \\
061006$^z$  & 0.4377     & $640_{-227}^{+144}$ & $2.0   \pm 0.3   $ \\
070714B     & 0.92       & $2150   \pm 1045  $ & $9.8   \pm 2.4   $ \\
080123      & 0.495      & $105    \pm 21.5  $ & $0.13  \pm 0.015 $ \\
080905A$^b$ & 0.122      & $579    \pm 77.5  $ & $0.032 \pm 0.003 $ \\
090426      & 2.609      & $177    \pm 72    $ & $5.4   \pm 0.65  $ \\
090510$^b$  & 0.903      & $8090   \pm 594   $ & $74.3  \pm 3.2   $ \\
100117A$^b$ & 0.92       & $549    \pm 84.5  $ & $0.81  \pm 0.10  $ \\
100206A$^b$ & 0.41       & $639    \pm 131   $ & $0.74  \pm 0.50  $ \\
100625A$^b$ & 0.452      & $701    \pm 115   $ & $0.75  \pm 0.03  $ \\
100816A     & 0.805      & $247    \pm 8.5   $ & $7.3   \pm 0.25  $ \\
101219A$^b$ & 0.718      & $842    \pm 155   $ & $4.9   \pm 0.7   $ \\
101224A$^b$ & 0.72       & $568    \pm 475   $ & $0.34  \pm 0.07  $ \\
110717A     & 0.92       & $937    \pm 297   $ & $1.87  \pm 0.23  $ \\
111117A$^b$ & 2.221      & $663    \pm 70    $ & $8.3   \pm 0.9   $ \\
130603B     & 0.356      & $895    \pm 136   $ & $2.12  \pm 0.23  $ \\
150120A     & 0.46       & $190    \pm 146   $ & $0.21  \pm 0.05  $ \\
160410A     & 1.717      & $4660   \pm 1445  $ & $40.0  \pm 4.00  $ \\
160624      & 0.483      & $1247   \pm 531   $ & $4.5   \pm 0.4   $ \\
160821B     & 0.16       & $98     \pm 22    $ & $0.12  \pm 0.015 $ \\
170428A     & 0.454      & $1428   \pm 573   $ & $2.1   \pm 0.45  $ \\
\hline\end{tabular}
\label{tab1}

{\small $^{z}$belongs to the Type I Gold Sample of Z09\\ $^{b}$has $f_\mathrm{NC}\geq 0.5$ in the sample of B13}
\end{table}



\subsection{Computation of the viewing angle dependent prompt emission}

\begin{figure*}
    \centering
    \includegraphics[width=\textwidth]{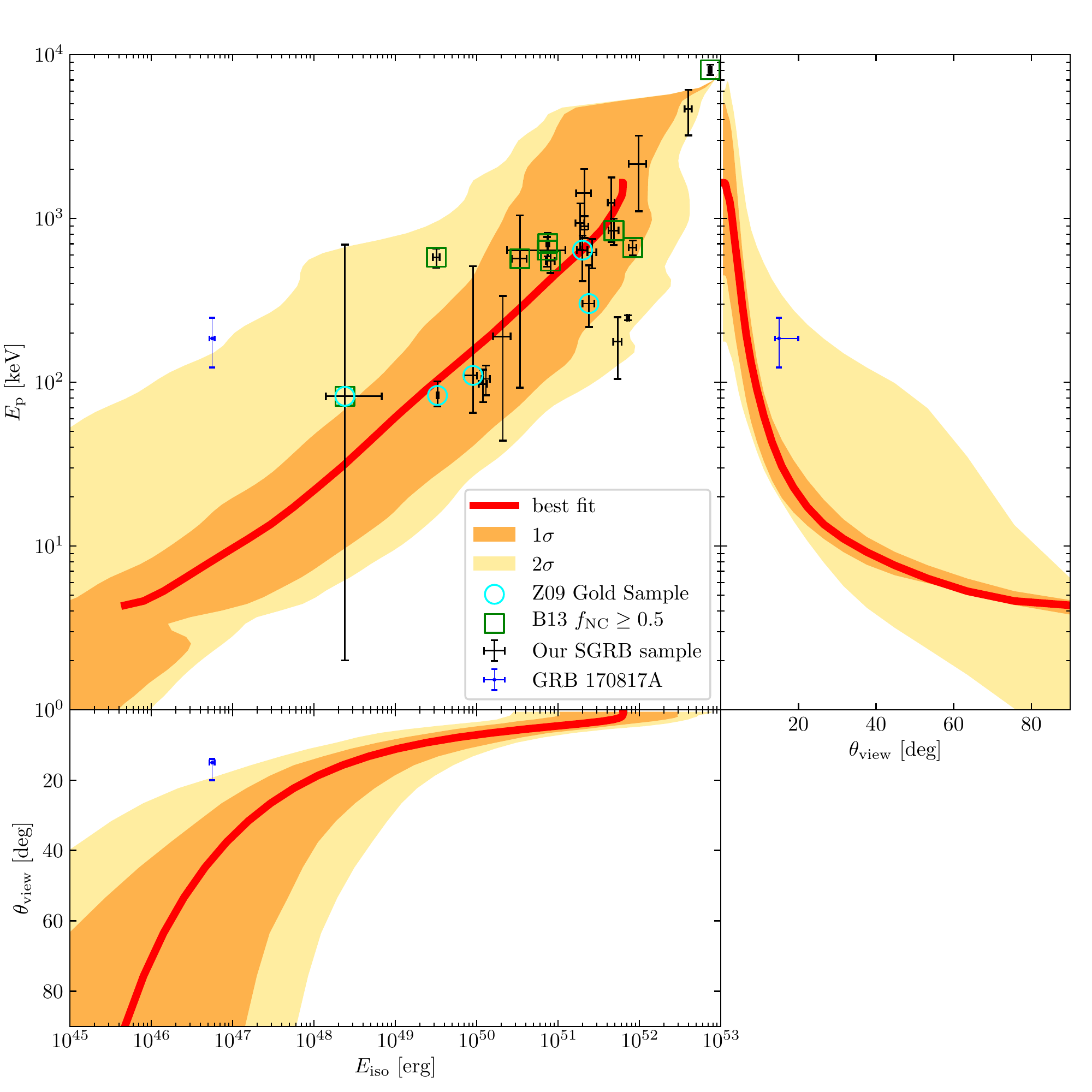}
    \caption{Possible prompt emission of a GRB 170817A-like jet compared to archival SGRB data. The bottom and right-hand panels show respectively the isotropic-equivalent energy $E_\mathrm{iso}(\theta_\mathrm{v})$ and the rest-frame SED peak photon energy $E_\mathrm{p}(\theta_\mathrm{v})$ as functions of the viewing angle $\theta_\mathrm{v}$ for the best fit jet structure of \cite{Ghirlanda2019} (red solid lines), along with their uncertainty regions ($1\sigma$ -- orange; $2\sigma$ -- yellow). In the top-left panel, the red solid line shows the corresponding track and uncertainty regions on the $(E_\mathrm{iso},E_\mathrm{p})$ plane. The plots are joint to ease the determination of the viewing angle that corresponds to a given point on the $(E_\mathrm{iso},E_\mathrm{p})$ track. The blue cross shows the observed properties of GRB 170817A, while black crosses show the SGRBs from our sample (\S\ref{sec:sample}). Cyan circles mark bursts which belong to the Type I Gold Sample in Z09, while green squares mark bursts that have $f_\mathrm{NC}\geq 0.5$ in the analysis of B13.}
    \label{fig:prompt}
\end{figure*}

We here attempt a comparison of the putative on-axis prompt emission properties of our GRB 170817A-like jet with the sample described in the preceding section. While there is a broad consensus within the community about the interpretation of GRB afterglows as due to synchrotron emission from non-thermal electrons accelerated at the external shock caused by the jet expansion in the surrounding medium \citep{Meszaros1993a}, the prompt emission mechanism remains elusive. The most common scenarios require partial conversion of the jet kinetic (e.g.~through internal shocks -- \citealt{Rees1994}) or magnetic (e.g~through reconnection -- \citealt{Thompson1994,Spruit2001}) energy to internal energy (typically in the form of a non-thermal population of electrons), which is then radiated in the $\gamma$-ray band through synchrotron, inverse Compton, or both \citep{Zhang2014}. Regardless the details of the emission process, several studies seem to indicate that the radiated energy typically amounts to $\sim 10$ -- $20\%$ of the jet kinetic energy (usually estimated by comparing the prompt $E_\mathrm{iso}$ with the jet kinetic energy derived from modelling of the afterglow, as in e.g.~\citealt{Fong2015,Beniamini2016}). We thus assume that a fraction $\epsilon=0.2$ of the kinetic energy in each jet solid angle element is radiated away in the form of high-energy photons, while $(1-\epsilon)$ remains in the form of kinetic energy that powers the afterglow. We assume the comoving emission to be isotropic, with an angle-independent comoving spectrum, so that the radiated energy per unit solid angle, per unit frequency, at observer frequency $\nu$ is given by 
\begin{equation}
    \frac{dE_\mathrm{\gamma}}{d\Omega d\nu}(\theta,\nu,\theta_\mathrm{v}) = \eta \frac{\delta^2(\theta,\phi,\theta_\mathrm{v})}{\Gamma(\theta)} \frac{dE}{d\Omega}(\theta) S((1+z)\nu/\delta)
    \label{eq:dE_dOmega_dnu}
\end{equation}
where $\eta=\epsilon/(1-\epsilon)$ and the comoving spectral shape $S(\nu')$ is normalised so that $\int_0^\infty S(\nu')d\nu'=1$ (primed quantities are in the jet comoving frame). 
We assume a cut-off power law comoving spectral shape, namely
\begin{equation}
    S(\nu')\propto \nu'^{a}\exp\left[-(1+a)\nu'/\nu'_\mathrm{p}\right]
    \label{eq:comoving_spectral_shape}
\end{equation}
With these assumptions, we can compute the prompt emission isotropic-equivalent energy $E_\mathrm{iso}(\theta_\mathrm{v})$ and the spectral energy distribution (SED) peak energy $E_\mathrm{p}(\theta_\mathrm{v})$ as measured by observers at different viewing angles, following \citet{Salafia2015}, that is
\begin{equation}
    \begin{split}E_\mathrm{iso}(\theta_\mathrm{v})=
        \int_{0}^{1}d\cos\theta\int_{0}^{2\pi}d\phi \int_0^\infty d\nu\,\frac{dE_\gamma}{d\Omega d\nu} = \\
        = \int_{0}^{1}d\cos\theta\int_{0}^{2\pi}d\phi \;\eta\frac{\delta^3(\theta,\phi,\theta_\mathrm{v})}{\Gamma(\theta)}\frac{dE}{d\Omega}(\theta)
    \end{split}
    \label{eq:Eiso(thv)}
\end{equation}
and
\begin{equation}
    E_\mathrm{p}(\theta_\mathrm{v})=h\times\mathrm{argmax}\left(\nu \frac{dE_\mathrm{\gamma}}{d\nu}\right)
    \label{eq:Ep(thv)}
\end{equation}
where $h$ is Planck's constant, and
\begin{equation}
    \frac{dE_\mathrm{\gamma}}{d\nu} = \int_{0}^{1}d\cos\theta\int_{0}^{2\pi}d\phi\, \frac{dE_\gamma}{d\Omega d\nu}
\end{equation}
Let us note that the comoving spectral shape $S(\nu')$ is needed only to compute the dependence of the spectral energy distribution (SED) peak energy $E_\mathrm{p}$ on the viewing angle, so its detailed shape is unimportant. We set the comoving peak photon energy to $h\nu'_\mathrm{p}=3\,\mathrm{keV}$ and the low-energy spectral index to $a=0.3$, which is typical for observed spectra of SGRB \citep{Nava2011}. The result is shown in Figure~\ref{fig:prompt}, where red solid lines refer to the central values of the best fit parameters, while the shaded regions stem from the uncertainty in the jet model parameters from \citet{Ghirlanda2019}. The regions are constructed by randomly selecting 1000 posterior samples from the Markov chain Monte Carlo of \citet{Ghirlanda2019} and computing the corresponding $E_\mathrm{iso}(\theta_\mathrm{v})$ and $E_\mathrm{p}(\theta_\mathrm{v})$. The shaded regions in Fig.~\ref{fig:prompt} show the areas containing the 68\% (labelled  ``$1\sigma$'') and 95\% (labelled  ``$2\sigma$'') of the resulting curves. 

The predicted energy and SED peak for an on-axis observer (top-right end of the red curve in the upper panel of Fig.~\ref{fig:prompt}) fall right in the middle of the known population (represented by black crosses in Fig.~\ref{fig:prompt}), and the lower energy SGRBs in our comparison sample seem to follow the curve that connects $E_\mathrm{iso}$ and $E_\mathrm{p}$ as seen by observers at different viewing angles. This is reminiscent of previous results \citep{Salafia2015} that showed that the Amati correlation \citep{Amati2002} can be interpreted as a viewing angle effect in the quasi-universal structured jet scenario. The highest $E_\mathrm{iso}$ and $E_\mathrm{p}$ in the sample are about an order of magnitude larger, but they are still consistent with the one sigma uncertainty in the structure, and they can easily be accommodated if some scatter in the quasi-universal structure properties is allowed. The values of $E_\mathrm{iso}$ and $E_\mathrm{p}$ predicted by our simple model for an observer at 15 -- 20 degrees (i.e.~the inferred viewing angle of GRB 170817A -- \citealt{Ghirlanda2019,Mooley2018}), on the other hand, are not consistent with the observations of GRB 170817A. In particular, the predicted $E_\mathrm{iso}$ is too high, while $E_\mathrm{p}$ is too low. 

Let us note that, on the other hand, our simplifying assumption of angle-independent efficiency and comoving spectrum is probably not realistic: all most popular prompt emission scenarios -- dissipation of jet energy by internal shocks \citep{Rees1994} or magnetic reconnection \citep[e.g.][]{Zhang2011,Lazarian2003}, happening either above or below \citep{Rees2005} the photosphere, followed by emission by synchrotron \citep{Ghisellini2000,Ravasio2018,Oganesyan2019} or photospheric radiation -- lead to some dependence of the typical photon energy on the jet luminosity and magnetisation \citep[for a comprehensive discussion, see][]{Zhang2002}.

Indeed, while the discrepancy in $E_\mathrm{iso}$ can be ``cured'' by introducing a steepening of the kinetic energy structure at large angles, e.g.~by assuming a Gaussian structure instead of a power-law one (Gaussian structures have been successfully used in fitting the GRB 170817A afterglow light curves -- see e.g.~\citealt{Troja2018,Hotokezaka2018a} -- and are compatible with the results of \citealt{Ghirlanda2019}, as shown in their Figure~S6), a possibly more natural approach would be to assume an angle-dependent efficiency $\epsilon=\epsilon(\theta)$ that decreases away from the jet axis, which might be easily justified in both the internal shock scenario (since the wings are slower on average, implying a necessarily lower contrast in relative Lorentz factors between subsequent ejection episodes) or in the magnetic reconnection scenario (where magnetisation in the wings could be lower due to entrainment of ambient material as the jet punches out of the merger ejecta). Let us note that a low prompt emission efficiency at large angles is supported by several observational arguments also in long GRBs \citep{Beniamini2019b}, and that it may be a requirement to explain the SGRB luminosity function within the structured jet scenario \citep{Beniamini2019a}. A lower magnetic field and/or a lower emission efficiency at large angles would also imply a less effective electron cooling: this in turn would lead the comoving typical photon energy $h\nu'_\mathrm{p}$ to increase with the angular distance from the jet axis, thus also solving the under-prediction of $E_\mathrm{p}$ (see e.g.~\citealt{Ioka2019} -- which circulated in pre-print form during the preparation of this work -- where the authors indeed assume the comoving peak photon energy to increase away from the jet axis in order to reproduce the observed quantities). An anti-correlation between the source luminosity and the peak of its SED is already seen in Blazars, where it can be explained in a similar manner \citep{Ghisellini1998,Ghisellini2017}.

Alternatively, the process that produced the GRB 170817A prompt emission could be different from that of its on-axis siblings. A possible candidate could be emission associated to the cocoon shock breakout \citep[see e.g.][]{Kasliwal2017,Gottlieb2017a}, which may dominate only for off-axis observers.

\section{Discussion}

\subsection{Plausibility of a quasi-universal jet structure in SGRBs}
The possibility of a quasi-universal jet structure in SGRBs is directly related to the degree of diversity in their progenitors. If neutron star binaries are the main progenitors of SGRBs, and if their component masses are narrowly distributed around typical values, then the merger outcome will be similar in most cases, thus leading to similar jets. On the other hand, relatively small variations in the component masses could lead to qualitatively different outcomes, such as different merger remnants (a black hole, a meta-stable proto-neutron star, a stable neutron star -- e.g.~\citealt{Bartos2013}), each potentially leading to very different jet properties. Still, it may be the case that a GRB jet is produced only if certain conditions are met, e.g.~not in the case of a direct collapse to a black hole neither in the case of a long-lived neutron star: this would again narrow down the range of properties of SGRB progenitors. The existence of a quasi-universal jet structure can be unveiled by joint detections of SGRBs and GWs during the next decade \citep{Beniamini2019a}, and will therefore provide powerful insights about their progenitors.

\subsection{How likely is the detection of an associated kilonova if the jet is observed on-axis?}

\begin{figure*}
    \centering
    \includegraphics[width=\textwidth]{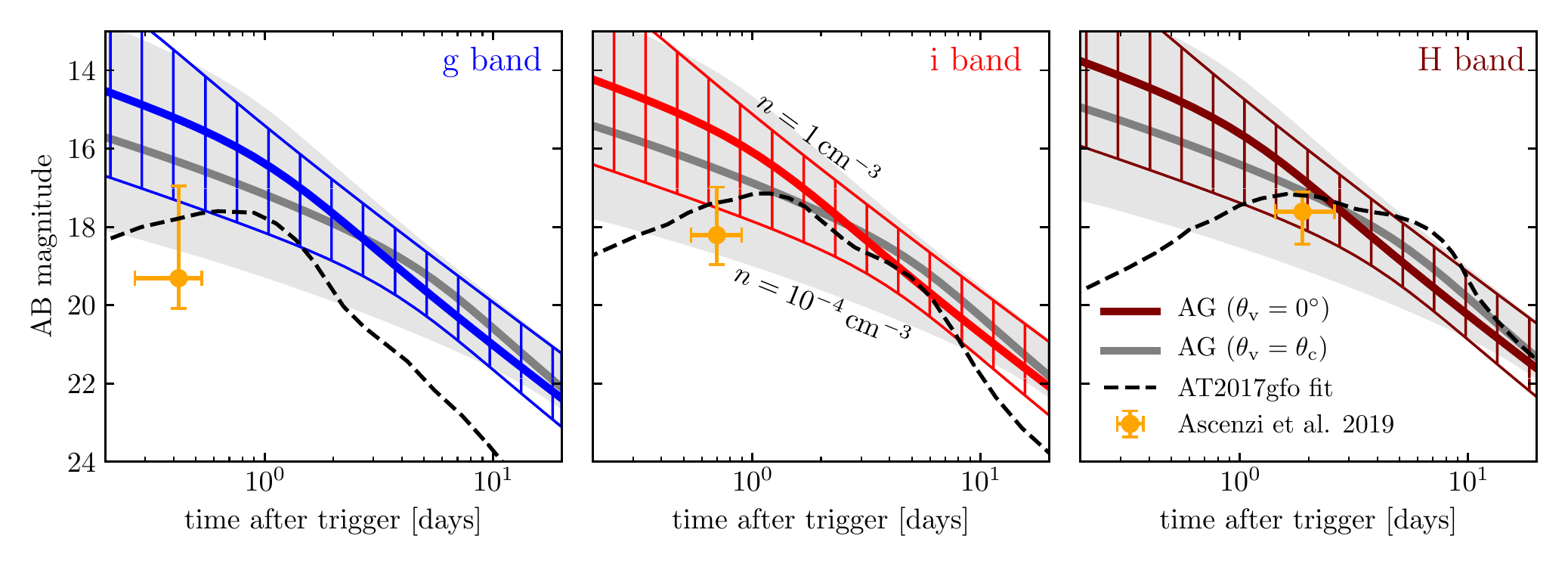}
    \caption{Comparison of the on-axis afterglow of a GRB 170817A-like jet with an AT2017gfo-like kilonova. Each panel shows light curves in a different band, as indicated in the upper right corner. Solid coloured lines represent the on-axis ($\theta_\mathrm{v}=0^\circ$) jet afterglow (with best fit parameters and $n=5\times 10^{-3}\,\mathrm{cm^{-3}}$), with the hatched region showing the portion of the plane span when assuming ISM densities between $n=10^{-4}\,\mathrm{cm^{-3}}$ and $n=1\,\mathrm{cm^{-3}}$. Grey solid lines and the grey shaded region show the same information for a jet seen at the core border ($\theta_\mathrm{v}=\theta_\mathrm{c}$). Black dashed lines show the AT2017gfo kilonova best fitting model from \cite{Villar2017}. The orange points show the median kilonova peak magnitudes and peak times in the three bands, as inferred by \citet{Ascenzi2019} (the error bars enclose 68\% of the events according to their distributions). }
    \label{fig:kn_comparison}
\end{figure*}

Several possible detections of kilonovae associated to known cosmological SGRBs have been claimed in the literature: GRB130603B \citep{Tanvir2013, Berger2013}, GRB050709 \citep{JinHotLi2016}, GRB060614 \citep{JinLiCano2015, YangJin2015} and GRB150101B \citep{TrojaRyanPiro2018, RossiStratta2019}\footnote{Recently \citet{RossiStratta2019} found further possible kilonova candidates in association with several other GRBs.}. 
\citet{Ascenzi2019} constructed a distribution of the peak absolute magnitude and peak time of kilonovae under the assumption that all these claims are correct, and using other non-detections as upper limits. 

Within the assumption of a quasi-universal structure, it is interesting to ask how often we should expect such a detection to be possible, given that the properties of the SGRB afterglow depend mainly on the extrinsic parameters. 
Since the KN and the GRB jet are located at the same distance, the only relevant extrinsic parameters are the viewing angle and the ISM density. 
For the usual SGRBs at cosmological distances the former is necessarily $\theta_\mathrm{v}\lesssim \theta_\mathrm{c}$. The latter, on the other hand, can vary by orders of magnitude: this is thus the dominant parameter for what concerns the SGRB. 

The kilonova emission is unaffected by the ISM density, and is also not affected significantly by relativistic beaming, given the 
small velocities involved ($\lesssim 0.1 c$). 
Viewing angle effects on the light curve could arise due to projection effects and to different optical depths to be crossed by photons emitted at different angles \citep[e.g.][]{Wollaeger2017}. These effects are still matter of debate, but they are expected to be small. For this reason, we take AT2017gfo as representative of a typical SGRB-associated kilonova seen on-axis, and we consider the distributions by \citet{Ascenzi2019} as an estimate of the range of variability of SGRB-kilonovae in general. We compute the afterglow light curves of our GRB 170817A-like jet as seen on-axis and just outside the core, varying the ISM number density between $10^{-4}\,\mathrm{cm^{-3}}$ and $1\,\mathrm{cm^{-3}}$, and keeping the luminosity distance fixed at $d_\mathrm{L}=40\,\mathrm{Mpc}$. The resulting afterglows span the hatched (on-axis jet) and grey shaded (off-core jet) regions in Fig.~\ref{fig:kn_comparison} when observed in the g, i and H bands. In each panel of the Figure, we also plot the AT2017gfo best fitting model from \citealt{Villar2017} (dashed black lines) and the medians of the distributions of peak times and peak magnitudes from \citealt{Ascenzi2019} (orange dots), where the error bars show the $1 \sigma$ scatter. This simple comparison suggests that  (in the quasi-universal jet hypothesis) an on-axis SGRB afterglow most often outshines the typical kilonova in the bluer bands, unless the progenitor neutron star merger takes place in a very low ISM density region. Infrared observations at few days post-trigger may be the most favourable option when looking for a kilonova signature associated to an SGRB afterglow. This is conistent with the fact that the first, and most solid, claim of such a detection (in GRB 130613B, \citealt{Tanvir2013}) is indeed due to an excess in the light curve at $\sim 9\,\mathrm{d}$ as seen in the WFC3/F160W filter of HST, whose central wavelength is\footnote{\url{http://www.stsci.edu/hst/wfc3}} $\sim 1500\,\mathrm{nm}$, i.e.~close to the H band.

\section{Conclusions}

Despite more than sixty years of observations and theoretical work, several questions about gamma-ray bursts remain unanswered. The diversity in their population, concerning energy, luminosity, duration, variability, afterglow decay rates, seems to contradict the narrow range of properties expected from their progenitors -- either neutron star mergers or pre-collapse Wolf-rayet stars. Starting about twenty years ago, several authors \citep[the first possibly being][]{Lipunov2001,Rossi2002,Zhang2002} suggested the possibility to trace back this diversity to the jet viewing angle, as in the unification scheme of active galactic nuclei \citep{Urry1995}. 

Our results, summarized in Figures~\ref{fig:afterglows} and \ref{fig:prompt}, show that GRB 170817A would appear as a standard SGRB if seen on-axis, providing support to this scenario.
The population of SGRB jets observed off-axis will soon increase in number, as the sensitivity of gravitational wave interferometers improves. We expect a large diversity in their properties, but due to extrinsic parameters only, such as
the viewing angle and circum-merger medium density.

\begin{acknowledgements}
We thank the anonymous referee for insightful comments which helped to improve the manuscript. We acknowledge the INAF-Prin 2017 (1.05.01.88.06) and the Italian Ministry for University and Research grant ``FIGARO'' (1.05.06.13) for support. We also acknowledge the implementing agreement ASI-INAF n.2017-14-H.0. SA acknowledges the GRAvitational Wave Inaf TeAm - GRAWITA (P.I. E. Brocato) for support.
\end{acknowledgements}

\footnotesize{
\bibliographystyle{aa}
\bibliography{references}
}

\end{document}